\documentclass[authoryear,preprint,review,12pt]{elsarticle}

\usepackage{graphicx}
\usepackage{longtable}
\usepackage{array}
\usepackage{lscape}  
\usepackage{adjustbox}
\usepackage{multirow}
\usepackage{tabularx}  
\usepackage{booktabs}  
\usepackage{makecell}  

\usepackage{amssymb}
\usepackage{amsmath}

\journal{ Computers in Biology and Medicine}

\begin{document}

\begin{frontmatter}



\title{Reinforcement Learning for Ultrasound Image Analysis: A Comprehensive Review of Advances and Applications}


\author[label1]{Maha Ezzelarab}
\author[label1]{Midhila Madhusoodanan}
\author[label1]{Shrimanti Ghosh}
\author[label1]{Geetika Vadali}
\author[label1]{Jacob Jaremko}
\author[label1]{Abhilash Hareendranathan}

\affiliation[label1]{organization={Department of Radiology \& Diagnostic Imaging, University of Alberta}, 
            addressline={}, 
            city={Edmonton},
            postcode={T6G 2R3}, 
            state={Alberta},
            country={Canada}}

\begin{abstract}
Over the last decade, the use of machine learning (ML) approaches in medicinal applications has increased manifold. Most of these approaches are based on deep learning, which aims to learn representations from grid data (like medical images). However, reinforcement learning (RL) applications in medicine are relatively less explored. Medical applications often involve a sequence of subtasks that form a diagnostic pipeline, and RL is uniquely suited to optimize over such sequential decision-making tasks.
Ultrasound (US) image analysis is a quintessential example of such a sequential decision-making task, where the raw signal captured by the US transducer undergoes a series of signal processing and image post-processing steps, generally leading to a diagnostic suggestion. The application of RL in US remains limited. Deep Reinforcement Learning (DRL), that  combines deep learning and RL, holds great promise in optimizing these pipelines by enabling intelligent and sequential decision-making.
This review paper surveys the applications of RL in US over the last decade. We provide a succinct overview of the theoretic framework of RL and its application in US image processing and review existing work in each aspect of the image analysis pipeline. A comprehensive search of Scopus filtered on relevance yielded 14 papers most relevant to this topic. These papers were further categorized based on their target applications—image classification, image segmentation, image enhancement, video summarization, and auto navigation and path planning. We also examined the type of RL approach used in each publication.
Finally, we discuss key areas in healthcare where DRL approaches in US could be used for sequential decision-making. We analyze the opportunities, challenges, and limitations, providing insights into the future potential of DRL in US image analysis.

\end{abstract}


\begin{highlights}
\item Surveys reinforcement learning applications in ultrasound over the last decade.
\item Categorizes RL studies in ultrasound by classification, segmentation, and path planning.
\item Explores RL's potential for optimizing sequential decision-making in ultrasound imaging.
\item Provides insights into challenges and opportunities of DRL in healthcare applications.
\item Highlights future directions for RL in ultrasound image analysis pipelines.
\end{highlights}


\end{frontmatter}



\section{Introduction}

Over the past decade, artificial intelligence (AI) has been increasingly used to solve medical image analysis tasks such as image segmentation, classification, object detection, and image enhancement. A vast majority of these applications use Deep Learning (DL) models that learn hierarchical representations from images or, in general, grid-based data. DL techniques are generally trained using supervised learning approaches in which the model minimizes a cost function based on manually labeled data. A fundamental limitation of DL models is that they are usually trained only for a single subtask, such as object detection, image classification, or semantic image segmentation. In medical image analysis, these tasks represent sequential steps in an image analysis pipeline. However, the larger goal is usually aimed at improving overall patient outcomes by taking optimal treatment decisions based on each of these subtasks. Deep learning models are not designed to be optimized over such sequences of subtasks.

Reinforcement learning (RL), on the other hand, is well-suited for these settings as it is based on training an agent to optimize a long-term reward based on a series of actions. Delayed reward scenarios are also very common in medical applications. For example, in tumor detection, the initial task is to delineate the boundaries of the organs of interest, and those of the tumor. This is usually followed by a closer evaluation of the regions and tumor staging. Finally, this image-based assessment is combined with other information about the patient such as age, previous medical history, and previous interventions to determine the optimal course of action. Such sequences of subtasks can be handled elegantly in an RL framework whereas integrating this into existing supervised learning settings is non-trivial.

Deep Reinforcement Learning (DRL) techniques aim to combine the benefits of both deep learning and RL. By embedding a deep neural network model within the RL framework, DRL retains the representational power of conventional deep learning models, while allowing RL-based sequential decision-making. DRL models have certain unique advantages over conventional deep learning approaches:
\begin{enumerate}
    \item They can optimize a reward metric over a sequence of tasks.
    \item They can optimize the reward, even when presented with non-differentiable loss functions.
\end{enumerate}

Despite these advantages, the use of DRL remains unexplored in the field of medical imaging in general and ultrasound (US) imaging in particular. US imaging involves a series of signal processing and image processing subtasks, starting with pre-processing of raw radio frequency (RF) data that is received by the individual transducer elements. RL agents could be used at various points of this signal processing pipeline for optimization. 

In this paper, we review some of the key research in this direction published over the last decade and review \textit{14} papers in depth. We start with a review of RL techniques, introduce basic terminologies, and set the context for understanding the RL frameworks commonly used in image processing tasks. We then provide a brief overview of the US signal processing pipeline, highlighting various aspects of it that are well-suited for applying RL techniques. The remaining sections of the paper explain the methodology used to select \textit{14} papers. We categorize these papers based on the type of US imaging, target organ, and the type of RL formulation.

\section{Ultrasound Signal Processing Pipeline}
\begin{figure}
    \centering
    \includegraphics[width=0.95\textwidth]{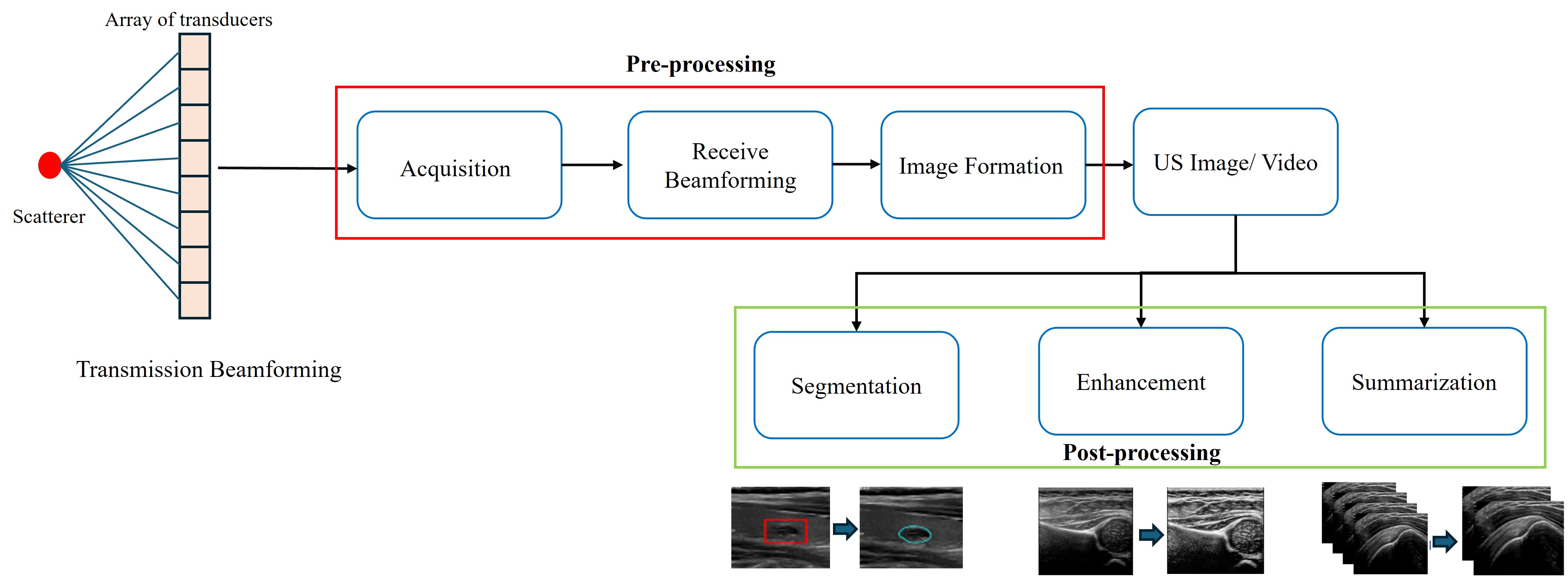}
    \caption{Ultrasound Imaging Pipeline}
    \label{fig:us_pipeline}
\end{figure}

In many clinical applications, US is used across a wide range of domains, including cardiology, obstetrics and gynecology, musculoskeletal imaging, pulmonology, and vascular imaging \cite{vedula2018}. US imaging involves a sequence of subtasks starting with the raw RF data generated from the transducer, which is processed to form the final B-mode image or video (as shown in Figure~\ref{fig:us_pipeline}). Most post-processing applications use this final image or video as input. However, the application of machine learning (ML) in earlier stages of the image analysis pipeline remains less explored.

Medical US operates at frequencies ranging from 1 to 50 MHz, depending on the required imaging depth for specific clinical applications \cite{szabo2013}. Lower frequencies (1–10 MHz) are typically employed for imaging deeper regions, such as the abdomen, gynecological and obstetric structures, and the breast, as they offer better penetration \cite{lockwood1996beyond}. Conversely, higher frequencies (10–50 MHz) are preferred for superficial imaging tasks, such as visualizing the skin, eyes, nerves, musculoskeletal structures, and performing intravascular imaging, where higher resolution is crucial \cite{szaboLewin2013}.

The US imaging process is performed through transmit and receive beamforming processes, which optimize the acoustic beam shape during transmission and reception. US imaging typically uses a pulse-echo approach, where an array of transducer elements emits an US beam. As the transmitted US wave passes through the tissue, the same transducer array captures the reflected signals over time, which are then used to reconstruct the US image \cite{szabo2013}.

The US imaging pipeline consists of four key stages: Transmission Beamforming, Acquisition, Receive Beamforming, and Image Formation. In the Transmission Beamforming stage, the number of transmissions and their beam profiles are selected based on the desired frame rate and image quality. The piezoelectric transducers are then configured to transmit the beams accordingly. After transmission, the transducer array receives echoes, which are demodulated and focused by applying time delays and phase rotations to generate the beamformed signal. This undergoes additional processing to correct artifacts (common in high frame-rate transmission modes) and apodization to reduce side lobes, which forms the Receive Beamforming stage. Finally, the complex signal is processed further by envelope extraction, log compression, and scan conversion to produce the final US image.

\section{Reinforcement Learning Framework}
RL is a type of ML where an agent learns to make sequential decisions by interacting with its environment. At its core, RL involves an agent, the model we aim to build, which interacts with an environment by observing its state and taking actions guided by a learned policy. The policy maps states to actions, serving as a stimulus-response rule. Following an action, the agent gets a reward, a scalar value that evaluates the quality of the agent's decision. Apart from this immediate reward, the agent also gets a delayed reward (otherwise called cumulative reward or credit assignment). This is a key aspect of the RL framework that differentiates it from Deep Learning.

\begin{figure}
    \centering
    \includegraphics[width=0.7\textwidth]{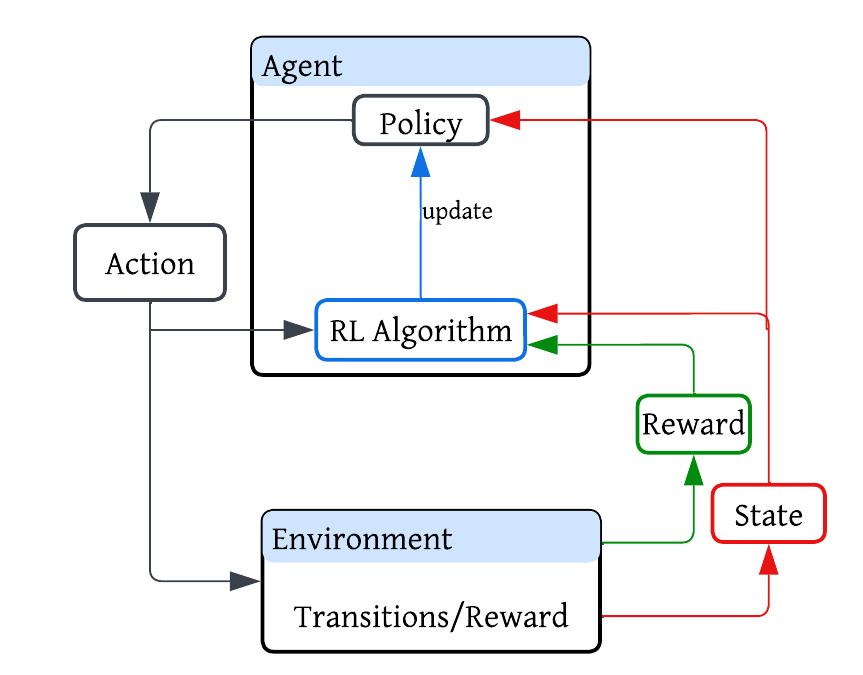}
    \caption{General Reinforcement Learning Framework}
    \label{fig:rl_framework}
\end{figure}

\subsection{Agent}
Key components of an RL agent include:
\begin{itemize}
    \item \textbf{Policy $\pi$:} A function that defines the agent's behavior, mapping states to actions.
    \item \textbf{Value Function:} Measures how good each state and/or action is.
    \item \textbf{Model:} A representation of the environment, consisting of:
    \begin{itemize}
        \item \textbf{Transition Model $p(s' = s_{t+1} \mid s_t = s, a_t = a)$:} Predicts the next state.
        \item \textbf{Reward Model $R$:} Predicts the immediate reward.
    \end{itemize}
\end{itemize}

The central concept in RL is maximizing the expected cumulative reward over time by selecting optimal actions. An RL environment can be described as a Markov Decision Process (MDP), a memoryless random process/sequence of random states. The states assume the Markov property whereby the current state contains all information needed to predict the next state, thus storing past states is unnecessary.

Unlike traditional decision-making, RL emphasizes not just the immediate reward but also how current actions influence future states, encapsulated in the concept of delayed reward. The cumulative future reward, or return, is defined as:
\begin{equation}
    G_t = R_{t+1} + \gamma R_{t+2} + \gamma^2 R_{t+3} + \dots
\end{equation}

Here, $\gamma$ ($0 \leq \gamma < 1$) is the discount factor, used to balance the trade-off between immediate and long-term rewards, avoid infinite returns, and reflect human preferences for immediacy. Here $t$ is the time step.

The value function quantifies the long-term benefit of states and actions under a policy $\pi$:
\begin{itemize}
    \item \textbf{State Value Function $v_{\pi}(s)$:} The expected return from state $s$ while following policy $\pi$:
    \begin{equation}
        v_{\pi}(s) = \mathbb{E}_{\pi}[G_t \mid S_t = s]
    \end{equation}
    \item \textbf{Action Value Function $q_{\pi}(s, a)$:} The expected return from taking action $a$ in state $s$, then following policy $\pi$:
    \begin{equation}
        q_{\pi}(s, a) = \mathbb{E}_{\pi}[G_t \mid S_t = s, A_t = a]
    \end{equation}
\end{itemize}

RL aims to solve the Bellman equations, which describe the value of a state or action in terms of immediate rewards and discounted future values. The Bellman expectation equations~\cite[eq. (3.14)]{sutton2018} define the state-value function $v_{\pi}(s)$ and action-value function $q_{\pi}(s, a)$ recursively, enabling iterative computation of optimal policies and value functions. The optimal value functions represent the maximum achievable performance under any policy. Solving the Bellman optimality equations~\cite[eq.(3.19 , 3.20)]{sutton2018} allows for determining the best actions to take in any state, effectively solving the MDP. These equations serve as the foundation for both model-based and model-free RL methods.

\subsection{Types of RL Approaches}
\begin{figure}
    \centering
    \includegraphics[width=0.90\textwidth]{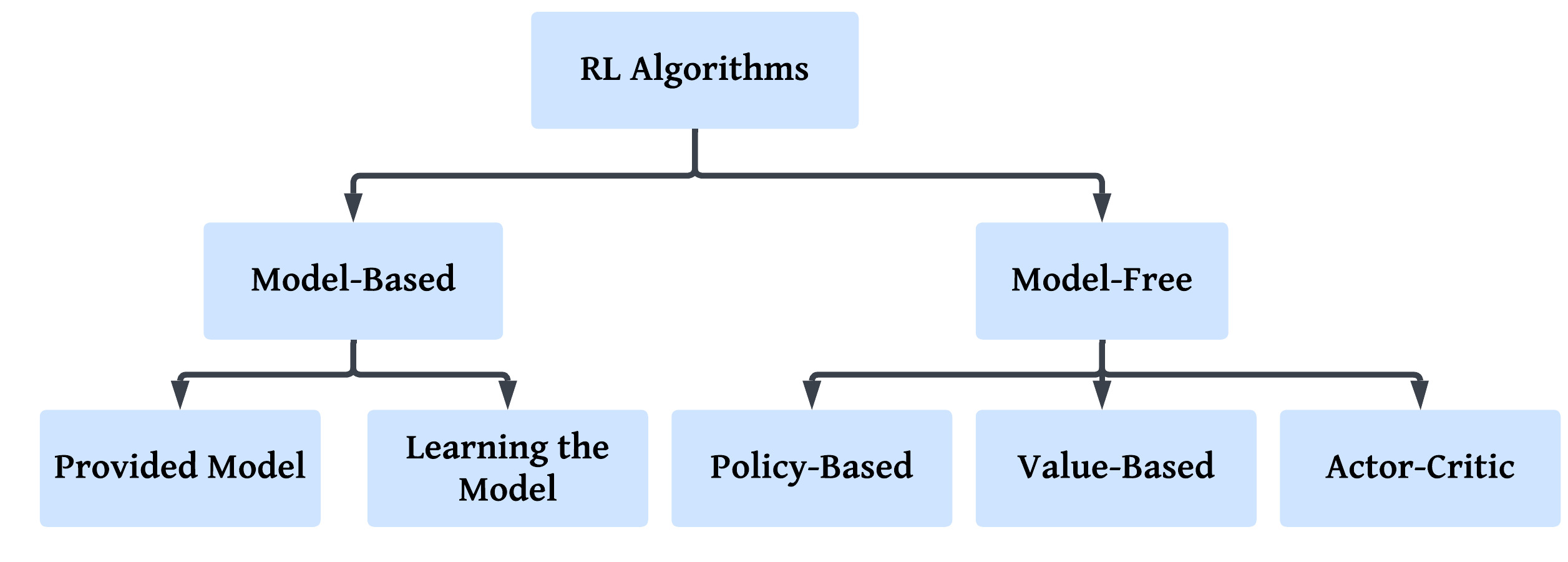}
    \caption{Types of Reinforcement Learning Approaches}
    \label{fig:rl_types}
\end{figure}

RL can be categorized into two main approaches based on how the agent interacts with and learns from the environment: Model-Free RL and Model-Based RL (refer Figure~\ref{fig:rl_types}).

\subsubsection{Model-Based RL}
This approach incorporates a representation of the environment, either provided or learned, to make decisions more efficiently by predicting outcomes of actions. It is split into:
\begin{itemize}
    \item \textbf{Provided Model:} The agent has access to a predefined environment model, including the transition probabilities $p(s' \mid s, a)$ and the reward function $R(s, a)$. This setup enables more straightforward planning. In this case, all variables in the Bellman expectation/optimality equations are known, and the MDP can be solved using Dynamic Programming techniques that decompose complex problems into subproblems. The two main methods for solving MDPs using dynamic programming are:
    \begin{enumerate}
        \item \textbf{Policy Iteration:} This approach aims to optimize the current policy to maximize the reward through:
        \begin{enumerate}
            \item \textit{Policy Evaluation:} Given a policy $\pi$, iteratively apply the Bellman expectation equation to calculate the value function $V_{\pi}(s)$ for each state $s$. This step continues until the value function converges.
            \item \textit{Policy Improvement:} Using greedy optimization, update the policy $\pi$ to maximize the value function $V_{\pi}(s)$. This involves selecting actions that maximize the expected return based on the current value estimates.
        \end{enumerate}
        \item \textbf{Value Iteration:} Combines policy evaluation and policy improvement into a single step. Instead of fully evaluating a policy, it directly updates the value of each state using the Bellman optimality equation. This process progressively adjusts value estimates toward their optimal values, skipping intermediate policy evaluations. Unlike Policy Iteration, Value Iteration does not explicitly maintain or evaluate a policy during the process. Hence, Value iteration is often faster in terms of iterations, though policy iteration may converge more steadily depending on the problem.
    \end{enumerate}
    \item \textbf{Learning the Model:} This approach involves two main stages:
    \begin{enumerate}
        \item \textit{Model Learning:} The agent first learns the transition dynamics and the reward function through interactions with the environment.
        \item \textit{Planning:} Once the model is learned, the agent uses this information to simulate potential future states and rewards. This enables the agent to perform decision-making by evaluating different action sequences before taking them in the real environment.
    \end{enumerate}
\end{itemize}

\subsubsection{Model-Free RL}
This approach focuses on directly learning optimal policies or value functions from interactions with the environment, without explicitly modeling the environment in case it is too complex or unknown. It is further divided into:
\begin{itemize}
    \item \textbf{Policy-based:} The agent learns a policy directly by optimizing it to maximize cumulative rewards. It is particularly effective in environments with continuous action spaces or when stochastic policies are required. The policy is typically updated using gradient-based optimization techniques, aiming to maximize the expected return.
    \item \textbf{Value-based:} This involves learning action-value functions to guide the policy indirectly by selecting actions that maximize the learned value function. It is highly efficient in discrete action spaces.
    \item \textbf{Actor-Critic:} Combines the strengths of both policy-based and value-based methods. The "actor" learns the policy, determining which action to take, while the "critic" evaluates the action by estimating its value using a value function.
\end{itemize}

Monte Carlo (MC) and Temporal Difference (TD) learning form the foundation of all model-free RL algorithms, differing in how they estimate value functions. MC relies on sampling complete episodes to calculate the average return for each state based on observed trajectories, making it suitable for episodic environments~\cite[Chapter 5]{sutton2018}. TD, on the other hand, updates value estimates incrementally after each time step, combining sampling with bootstrapping, which allows it to handle both episodic and continuous environments~\cite[Chapter 6]{sutton2018}. While MC is intuitive, TD’s flexibility makes it more widely applicable. Both approaches support on-policy and off-policy learning, enabling agents to learn from their own actions or from a separate behavior policy.

\subsection{Function Approximation and DRL}
All the previously discussed methods were initially designed for tabular settings, where the value function is represented using a lookup table:
\begin{itemize}
    \item Each state $s$ has an entry $V(s)$.
    \item Or, each state-action pair $(s, a)$ has an entry $Q(s, a)$.
\end{itemize}

However, in environments with large or continuous state/action spaces, tabular methods become infeasible due to memory constraints and the inability to generalize to unseen states. To scale up model-free methods, we replace the lookup table with a function approximator, such as a neural network, to represent the value function:
\begin{itemize}
    \item Instead of learning $V(s)$ for each state individually, we learn a function that maps $s$ to $v_{\pi}(s)$.
\end{itemize}
This reduces memory requirements and allows generalization to new, unseen states. This shift from tabular representations to function approximation lays the foundation for DRL methods.

\section{Methods}
\begin{figure}
    \centering
    \includegraphics[width=1.0\textwidth]{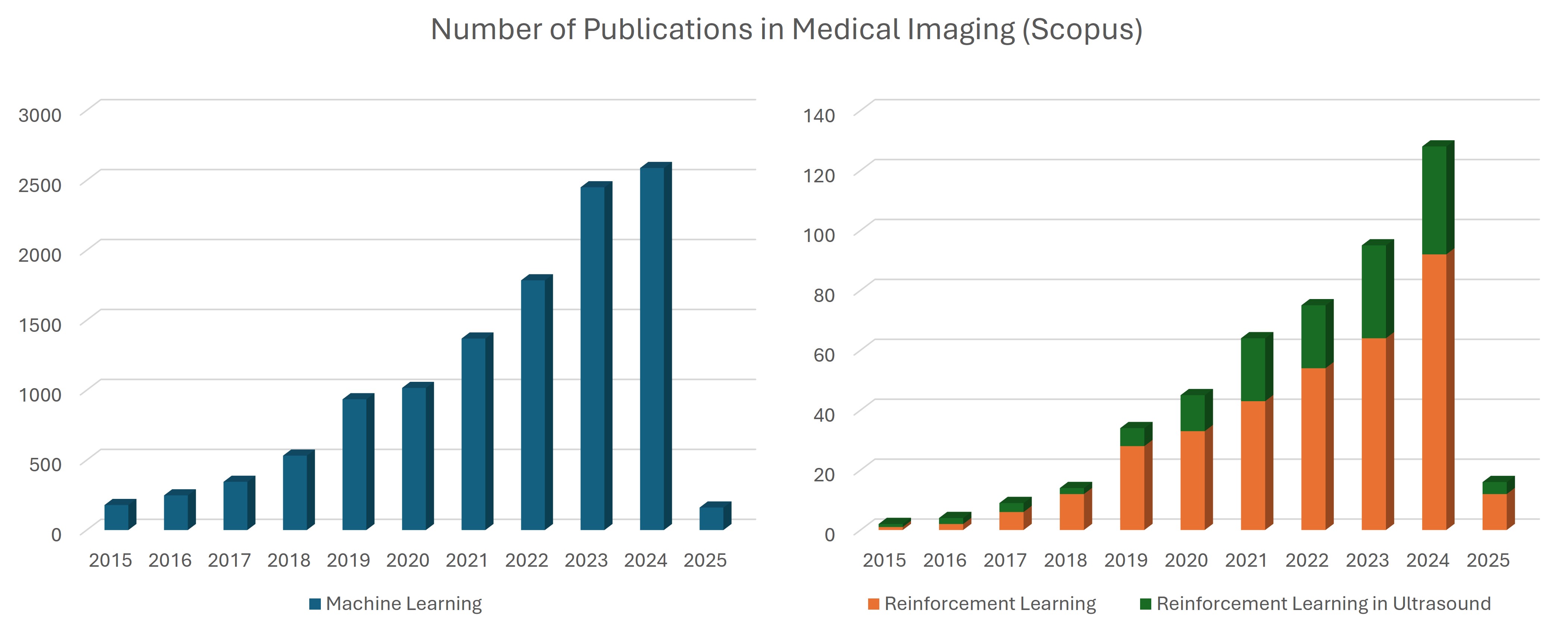}
    \caption{Number of publications related to machine learning (ML), reinforcement learning (RL), and RL applications in ultrasound.}
    \label{fig:statistical_analysis}
\end{figure}

We conducted a search using the Scopus database to identify publications related to ML, RL, and specifically, RL applications in US. The search targeted titles, abstracts, and keywords, utilizing the following queries:
\begin{itemize}
    \item "Machine Learning" AND ("Medical Imaging" OR "Medical Image Analysis")
    \item "Reinforcement Learning" AND ("Medical Imaging" OR "Medical Image Analysis")
    \item "Reinforcement learning" AND "ultrasound"
\end{itemize}
This search yielded 11,606 publications in ML and 347 publications in RL in the last decade, with an increasing number of publications each year, as shown in Figure~\ref{fig:statistical_analysis}. The search for RL in US yielded 139 publications. In addition, 2 papers were selected from a previous review paper authored by \cite{zhou2021} . After thorough evaluation, a total of 14 publications were selected for detailed review by the authors ME, MM, and SG.
Three reviewers (ME, MM, and SG) independently reviewed all publications, and any conflicts in categorization were resolved through consensus among all authors (ME, MM, SG, and AH). The reviews were performed by two graduate students (ME and MM) and one postdoctoral researcher (SG).

\section{Results}
Out of the 14 papers selected for review, 2 papers applied DRL for segmentation, 1 for classification, 3 for video summarization, 4 for enhancement, and 4 for path planning and autonomous navigation. Both policy-based and value-based approaches were utilized in these studies. Since most of these applications involve images, the Deep Q-Network (DQN) model, which uses a neural network to learn data representations, was applied in 3 applications, while the REINFORCE RL framework was used in 7 applications. These categorizations are summarized in Table~\ref{tab:publication_summary}.

\clearpage


\footnotesize{

\begin{longtable}{|p{2.2cm}|p{1.3cm}|p{2.1cm}|p{2.0cm}|p{1.4cm}|p{1.9cm}|}

    \caption{\centering{Reinforcement Learning Applications in Ultrasound Image Analysis}} 
    \label{tab:publication_summary} 
    \\
    \hline
    \makecell{Task \\ Category} &
    Authors & 
    \makecell{RL \\ Algorithm} &
    \makecell{Application/ \\ ROI} & 
    \makecell{Dataset/ \\ Code} &
    \makecell{Dataset \\ Size} \\ 
    \hline
    \endfirsthead 

    \hline
    \makecell{Task \\ Category} &
    Authors & 
    \makecell{RL \\ Algorithm} &
    \makecell{Application \\ /ROI} & 
    \makecell{Dataset/ \\ Code} &
    \makecell{Dataset \\ Size} \\ 
    \hline
    \endhead 

    \hline
    \multicolumn{6}{|r|}{{Continued on the next page}} \\
    \hline
    \endfoot 

    \hline
    \endlastfoot 
    Classification & Wang J et al. 2020 & REINFORCE & Breast cancer classification & Private / No & 1616 multimodal US image sets. \\
    \hline
    
    \multirow{2}{*}{Segmentation} & Yang H et al. 2020 & DQN & Catheter segmentation & Private / No & 88 3D US. \\ 
    \cline{2-6} 
    & Sahba et al. 2006 & REINFORCE & Soft Tissue / Prostate Gland & Private / No & - \\ 
    \hline  
    
    \multirow{2}{*}{Summarization} 
        & Liu T et al. 2020 & REINFORCE & Fetal US videos & Private / No & 50 videos. Key diagnostic planes labeled. \\ \cline{2-6}
        & Mathews et al. 2022 & Unsupervised RL using Convolutional Autoencoder Transformer Decoder & Lung US videos & Private / No & 100 Lung US videos. \\ \cline{2-6}
        & Huang R et al. 2022 & REINFORCE & Extract keyframes from breast US videos & Private / No & 2606 breast US videos. \\ \hline
    
    \multirow{4}{*}{Enhancement} 
        & Sun et al. 2024 & DQN & Underwater US & Public / Yes & 950 (UIEB Datasets). \\ \cline{2-6}
        & Jarosik et al. 2021 & REINFORCE & Breast US Denoising & Public / No & 5202 from BSD68 and Waterloo. 100 from OASBUD . \\ \cline{2-6}
        & Malamal et al. 2023 & REINFORCE & US Guided Needle Tracking & Private / No & - \\ \cline{2-6}
        & Stevens et al. 2022 & REINFORCE & Intravascular US imaging & Private / No & - \\ \hline
    
    \multirow{4}{*}{\makecell{Auto \\ Navigation \\ and \\Path Planning}} 
        & Yao et al. 2024 & BRLfD & Robotic Breast US & Private / No & - \\ \cline{2-6}
        & Li et al. 2021 & DQN & Musculo-skeletal & Private / No & 41 3D US. \\ \cline{2-6}
        & Amadou et al. 2023 & GCRL & TEE Imaging & Private / No & Synthetic. \\ \cline{2-6}
        & Ao et al. 2025 & PPO & Pedicle Screw Placement & Private but available on request / Yes & 25 vertebrae from ITIS ViP, 25 vertebrae phantoms from water model set, 35 vertebrae from real human US images. \\ \hline

\end{longtable}
}

\normalsize

\subsection{Classification using RL}

Deep learning approaches have been successfully used for various image classification applications in both natural images and medical images. Some of these approaches have also been applied to US image classification. However, image classification approaches using DRL in medical imaging are limited, with notable exceptions being in \cite{wang2020}.
Among the papers reviewed in this study, \cite{wang2020} applied RL for US image classification. Their objective was to combine information from four US modalities—B-mode, Doppler, Shear Wave Elastography, and Strain Elastography—to classify breast tumors as malignant or benign. Using conventional deep learning approaches,\cite{wang2020} trained a ResNet-18 model for each of the four imaging modalities. They then employed a reinforcement learning (RL) agent to optimally combine the classification results from the four ResNet models. The RL agent explored various weight configurations for the classification outcomes, with the aim of improving overall performance. Validation performance feedback was used as a reward signal, allowing the RL agent to adjust its parameters through the REINFORCE algorithm to maximize the classification accuracy.

\subsection{Segmentation using RL}
RL has emerged as a promising approach for medical image segmentation, addressing challenges like noise, variability, and complex structures by modeling segmentation as a sequential decision-making process. Early works, such as \cite{Sahba2006}, introduced an innovative approach to medical image segmentation through RL. The authors treat the segmentation task as a decision-making problem, where the model interacts with the image data to incrementally improve its segmentation results based on a reward mechanism. This RL-based framework allows the system to adapt to various challenges in medical imaging, such as variations in image quality, noise, and anatomical complexity. The key advantage of this approach lies in its ability to learn optimal segmentation strategies over time, improving both accuracy and robustness compared to traditional methods. The framework is particularly effective for biomedical and ultrasonic imaging applications, where precise and reliable segmentation is crucial for accurate diagnostics.
\cite{yang2020} advanced the application of RL by proposing a novel framework for catheter segmentation in 3D US imaging, tailored for cardiac interventions such as radiofrequency ablation. The method integrates DQN-based RL as a coarse localization step, addressing challenges like limited training data, high annotation costs, and class imbalance. The DQN agent navigates through 3D patches of the US image, iteratively adjusting the patch location via discrete actions guided by a reward function based on the Euclidean distance to the catheter center. This RL-guided localization reduces dependency on voxel-level annotations, requiring only region-level labels. The framework achieved a mean Euclidean localization error of 4.3 voxels in a 128³ voxel environment, demonstrating superior accuracy and efficiency compared to traditional supervised methods. This study highlights the potential of RL to enhance segmentation performance in 3D US imaging, particularly when annotation resources are limited.

\subsection{Summarization using RL}
US videos are generally long, containing 200–600 frames depending on the length of the sweep. These videos are either acquired using a 3D US transducer or by manually sweeping a 2D US transducer over the target areas. They often contain several redundant frames that do not contain clinically relevant information. The diagnostically useful information details can be captured in a small subset of these frames. 
As a result, video summarization techniques, which aim to identify the optimal subset of non-redundant frames that capture all essential information from a video, have gained significant research interest. In a Deep Learning framework, this can be formulated as a regression problem, where a neural network model is trained to select an optimal subset of frames based on manual annotations. However, these annotations often vary significantly even among experts, as the judgment of relevant frames can be subjective.
Video summarization can be approached as a sequence of subtasks within a DRL framework. For example, \cite{liu2020} developed a fully automatic DRL video summarization method to retain essential diagnostic information while reducing redundancy in US videos. Their approach utilized an encoder-decoder model, where the encoder was pre-trained on US standard plane detection annotations for feature extraction. The decoder was implemented as a Bi-LSTM, which modeled temporal dependencies and generated frame importance scores based on both past and future context. They used the REINFORCE algorithm to iteratively adjust the policy by selecting frames that maximize rewards for representativeness, diversity, and diagnostic quality.

\cite{huang2022} proposed KE-BUV, a reinforcement learning-based framework for automatic keyframe extraction in breast ultrasound videos, aimed at improving lesion diagnosis. The framework addresses challenges such as frame redundancy, variability in lesion characteristics (e.g., size, shape, location), and the integration of diagnostic knowledge into keyframe selection. KE-BUV incorporates a Detection-Based Nodule Filtering (DNF) module to preprocess videos and focus on lesion regions. A reinforcement learning agent, comprising a C3D feature extractor and Bi-LSTM, captures spatial and temporal features to sequentially select diagnostically relevant keyframes. The agent is guided by a customized reward mechanism, which integrates alignment with manual annotations, detection accuracy, and diagnostic relevance. The model was trained using the REINFORCE algorithm to optimize keyframe selection based on diagnostic utility. Additionally, the Attribute Classification Network (ACN) employs a group-aware focal loss to address class imbalance in malignancy indicator classification. KE-BUV achieved state-of-the-art performance, outperforming existing methods in alignment with expert annotations and diagnostic accuracy, while demonstrating robustness to varying video lengths and lesion characteristics.

\cite{mathews2022} presented a novel unsupervised reinforcement learning (RL) framework for ultrasound video summarization, addressing the challenges of long, redundant, and noisy medical scan videos. Their approach leverages a Convolutional Autoencoder and a Transformer decoder to extract diagnostically relevant frames without the need for manually annotated labels, making it highly scalable for real-world clinical applications. This fully unsupervised learning scheme was validated on lung ultrasound videos, achieving a precision of 64.36\% and an F1 score of 35.87\%, with an impressive video compression rate of 78\%. The method significantly reduces the data burden for emergency triage and telemedicine applications, allowing for faster and more efficient video analysis.

\subsection{Autonomous Navigation using RL}
The application of RL to US navigation has shown significant progress in recent years, with each contribution pushing the boundaries of autonomous ultrasound-guided interventions.
In \cite{li2021}, the authors proposed a DQN framework, addressing the challenge of autonomously controlling the 6D pose of an US probe for navigating toward standard scan planes. Their work was focused on enhancing image quality using a confidence-based approach within RL, optimizing scan plane navigation based on real-time feedback. The DQN estimates expected rewards through Q-values, making it suitable for the task of US probe positioning. The novel aspect of their work lies in integrating this DQN-based control with image quality improvement, enabling the probe to autonomously reach desired scan planes, despite the challenges posed by the complexity and variability of US images.

In \cite{amadou2024}, the authors took a significant leap forward with the introduction of Goal-Conditioned Reinforcement Learning (GCRL), a novel methodology for US navigation that goes beyond the ability to reach standard views. Their framework enables the probe to autonomously navigate to arbitrary, goal-specific views, making it a versatile tool for both diagnostic and interventional applications. What makes this work unique is the ability to target non-standard views, such as those required for left atrial appendage (LAA) closure, which had not been previously achievable using standard navigation methods. To enhance generalization across diverse patient data, \cite{amadou2024} introduced contrastive patient batching, which improves the critic's learning by sampling harder contrastive pairs to prevent overfitting. Additionally, they incorporated contrastive data augmentation loss to further improve the robustness and quality of the learned representations. Their model demonstrated competitive performance in navigating to both standard and non-standard views, showing promising results for real-world clinical procedures.

\cite{yao2024} proposed a decision-making framework for robotic breast US imaging, combining RL with learning from demonstration (LfD). The algorithm aims to guide the robotic system in acquiring high-quality US images by learning from expert demonstrations, which improves both learning efficiency and robustness. By leveraging expert-provided demonstrations, the robot can better adapt to various patient anatomies and diverse imaging conditions, making the imaging process more flexible and reliable. The RL component optimizes decision-making, allowing the robot to autonomously adjust probe positioning and movement to achieve the best image quality, thus reducing the reliance on manual input. The approach is shown to significantly improve imaging quality and efficiency, paving the way for autonomous and precise US systems that could revolutionize clinical practice, making breast imaging more accessible and consistent.

Recently, \cite{ao2025saferplan} introduced SafeRPlan, a safety-aware DRL framework for robotic spine surgery that autonomously plans safe pedicle screw trajectories using intraoperative US dorsal surface reconstruction. The core of SafeRPlan is using a Safe teacher-student reinforcement learning with distance-based safety filter involving an actor-critic network, where the actor generates actions (screw trajectory decisions) and the critic evaluates the chosen actions with a distance-based safety filter (DSF), which ensures safety by evaluating the predicted distance between the screw trajectory and vital structures. The safety filter operates in real-time during policy deployment, encouraging the agent to take safe actions when necessary, such as moving backward to avoid breaching bone. SafeRPlan also employs a teacher-student network, where the teacher agent is trained in a simulation environment using a constructed dorsal surface from preoperative CT/MRI data, providing optimal policies for safe task execution. The student agent learns from the teacher and is trained with domain randomization, which introduces noise and disturbances to simulate real-world conditions, thereby reducing the simulation-to-real gap. SafeRPlan demonstrates improvement in safety compared to baseline methods, balancing safety and stability while outperforming traditional registration-based methods. However, high-quality US surface reconstructions remain essential for achieving optimal surgical outcomes. 

These advancements collectively demonstrate the potential of RL to transform US imaging into a more flexible, accurate, and autonomous system for both diagnostic and interventional procedures.

\subsection{Image Enhancement using RL}
RL has been used to enhance image quality both in medical images and natural images. \cite{jarosik2021} introduced a pixel-wise RL approach for denoising US images, specifically targeting speckle noise. The method, known as pixelRL, employs predefined filters and pixel-level actions that are iteratively applied to improve image quality, measured by metrics such as peak signal-to-noise ratio (PSNR), which was increased from 14 dB to 26 dB while retaining the structural information in the original image, as indicated by an increase in the structural similarity index (SSIM) from 0.22 to 0.54. 

In the field of intravascular ultrasound (IVUS) imaging, \cite{stevens2022} proposed a deep RL framework designed to accelerate image acquisition by optimizing sampling patterns. By intelligently adapting sampling strategies during data acquisition, the method reduces image acquisition time while maintaining high image quality, thereby enhancing real-time vascular diagnostics. This approach highlights the ability of RL to optimize imaging processes for speed and accuracy.

Malamal and Panicker (2023) addressed the challenge of needle tracking in ultrasound-guided interventions by integrating unsupervised RL with Accelerated Adaptive Minimum Variance Beamforming. Their method adapts beamforming parameters in real-time to enhance image quality and tracking accuracy without requiring labeled data, offering a computationally efficient solution for real-time guidance of needle placement in clinical settings. This RL-based approach improves spatial resolution and minimizes noise, making it a significant advancement for ultrasound-guided procedures.

\cite{sun2022} introduced an RL framework for enhancing underwater images, which are often degraded by color distortion, low contrast, and noise due to environmental factors like light absorption and scattering. Using DQN and MDP, their method iteratively selects optimal enhancement actions tailored to each image's unique characteristics, significantly improving color and contrast. This RL-based approach outperforms traditional enhancement methods, providing a scalable solution for underwater imaging and exploration.

\section{Discussion}

Ultrasound image analysis involves a series of sequential tasks that make it uniquely suited to be modeled as an RL problem. In most cases, there is also a delay in reward or credit assignment, where the reward from a certain decision or sub-action becomes available only at a much later stage in the image analysis pipeline. For instance, beamforming techniques applied to raw US signals inherently affect the quality of the US image produced for subsequent assessment and diagnosis. However, these rewards are evident only after the original image has been analyzed using various post-processing routines.

The potential of using DRL in US image analysis has not yet been fully explored, with only a few publications in this area, which is significantly fewer than other ML-based approaches for medical image analysis. We categorized these applications into segmentation, classification, summarization, navigation, and image enhancement. In this section, we analyze other aspects of these works, such as the DRL formulation used (e.g., REINFORCE vs. DQN), dataset size and availability, key challenges, and potential areas where RL could be applied in the coming years.

\subsection{DRL Formulations and Approaches Used}

DRL builds on the foundational principles of RL by integrating deep neural networks as function approximators. This innovation enables DRL to tackle high-dimensional and continuous action policies directly from raw data, such as pixel intensities in US images. Two prominent DRL algorithms frequently encountered are REINFORCE and DQN.

REINFORCE is a policy-based method that directly optimizes the policy by maximizing expected cumulative rewards using policy gradient techniques. It is particularly effective in scenarios with continuous or high-dimensional action spaces. Out of the papers surveyed here, 7/14 papers used the policy-based REINFORCE approach for image classification, segmentation, summarization and image enhancement.

Another popular model used in RL approaches is the DQN, which is a value-based method that combines Q-learning with deep neural networks to approximate the action-value function. DQN has demonstrated effectiveness in discrete action spaces. It combines the ability of CNN models to learn a hierarchical representation of grid data with the flexibility of an RL framework. 3/14 papers used DQN-based approaches for image enhancement, segmentation, and path planning.

These methods form the backbone of DRL approaches in US image analysis. For readers interested in the theoretical and mathematical foundations of these algorithms, a detailed explanation is provided in Appendix A at the end of this paper.

\subsection{Dataset Size and Availability}

The reviewed studies highlight a wide range of data sizes and availability, with most relying on private datasets, which limits reproducibility. Dataset sizes vary significantly, from 50 annotated US videos \cite{liu2020} to over 2,600 breast US videos \cite{huang2022}. Public datasets were referenced in only a few studies, such as \cite{jarosik2021}, who used BSD68 and Waterloo datasets for training and OASBUD for evaluation. In contrast, many studies, including \cite{yao2024} and \cite{amadou2024}, relied on private or synthetic data, often without disclosing detailed sizes (see Table~\ref{tab:publication_summary}). Only two studies, \cite{sun2022} and \cite{ao2025saferplan}, provided their code, with \cite{ao2025saferplan} also offering three datasets upon request.

Overall, the scarcity of publicly available datasets, accessible code, and detailed methodological transparency highlights a critical limitation, underscoring the need for more open resources to promote reproducibility and wider adoption of RL methods in US imaging.

\subsection{Challenges and Limitations}

There are challenges in adapting RL techniques for medical imaging use cases, including US. Some of these challenges are common to ML approaches in general, such as the scarcity of labeled data and variability in ground truth annotations. ML models are generally trained using supervised learning approaches and require a large number of labeled or annotated datasets.

In natural image datasets, annotations are often generated through crowdsourcing, where users label large numbers of images over the internet. However, this approach is not feasible for medical images due to data privacy constraints and the need for specific expertise in labeling. As a result, medical image databases are typically much smaller compared to natural image databases. Even in cases where ground truth labels are obtained, there can be variability in annotations, even among experts. Similarly, deep learning and reinforcement learning models generally involve the optimization of millions of trainable parameters, which requires a significant amount of time (potentially several days) and computational resources.

The RL framework also has unique challenges. The performance of an RL model is heavily dependent on how the state, actions, and rewards are defined. Different formulations can drastically change the performance, even when applied to identical datasets. Another key limitation is that there are only a few RL studies with publicly available code, making it difficult to reproduce some of the results reported in the literature.

\subsection{Importance or Potential of RL in Healthcare}

Despite these challenges, RL holds immense potential in healthcare applications. Specifically, RL approaches are well-suited for sequential tasks that involve delayed rewards. This scenario arises quite often in medical US. For instance, RL can automate image acquisition by guiding US probe positioning in real-time, optimizing the position based on the final diagnostic utility of the acquired image.

It could also potentially be used to optimize treatment planning, improve clinical workflows, and deliver personalized care based on imaging data and patient demographics. RL can also be combined with existing deep learning approaches, leveraging the ability of deep learning models to represent complex image features. Applications of DRL, such as network architecture search to identify the optimal deep learning model for a given dataset, are yet to be fully explored in the context of US data.

\section{Conclusion}

In this review, we examined the application of reinforcement learning (RL) in ultrasound imaging over the past decade. While RL remains underexplored in this field, its ability to model sequential decision-making makes it well-suited for tasks such as image acquisition, classification, segmentation, video summarization, enhancement, and path planning. Ultrasound’s affordability, portability, and real-time imaging are offset by challenges like operator dependency, image quality issues, and motion artifacts. RL has shown potential in addressing these limitations through robotic navigation, dynamic image enhancement, and adaptive segmentation. However, barriers such as the need for annotated datasets, realistic simulations, and real-time performance optimization persist. Advancing RL in ultrasound requires interdisciplinary collaboration, data-efficient techniques, and clinically relevant frameworks to unlock its impactful potential in US medical imaging.

\appendix
\section{Appendix A}

\subsection{2.3.1. Preliminary}

\subsubsection{2.3.1.a. REINFORCE}
The REINFORCE algorithm is a fundamental policy gradient method used in DRL to optimize policies in RL problems. It directly parameterizes the policy and employs gradient-based optimization to maximize the expected cumulative reward. The objective is to maximize the expected return, $R$, defined as the expectation of cumulative rewards over trajectories. Using the policy gradient theorem, the gradient of the objective function is expressed as the expected product of the gradient of the log-probability of the policy and the return from the current state. The policy parameters are updated iteratively using the rule:
\begin{equation}
\theta \leftarrow \theta + \alpha \nabla_\theta \log \pi_\theta(s_t, a_t) G_t
\end{equation}
where $\alpha$ is the learning rate, $\nabla_\theta \log \pi_\theta(s_t, a_t)$ is the gradient of the log-probability of the action given state $s_t$, and $G_t$ is the cumulative return from time step $t$. Despite its simplicity and effectiveness in environments with continuous action spaces, REINFORCE suffers from high variance in gradient estimates, which can lead to unstable training. Its performance also depends heavily on the choice of learning rate and reward normalization.

\textbf{Algorithm:}
\begin{itemize}
    \item Initialize policy parameters arbitrarily.
    \item For each episode:
    \begin{itemize}
        \item Generate a trajectory by following the policy $\pi$.
        \item For each time step in the trajectory:
        \begin{itemize}
            \item Compute the return $G_t$.
            \item Update the policy parameters: $\theta \leftarrow \theta + \alpha \nabla_\theta \log \pi_\theta(s_t, a_t) G_t$.
        \end{itemize}
    \end{itemize}
    \item Repeat until convergence.
\end{itemize}

\subsubsection{2.3.1.b. DQN}
Deep Q-Networks (DQN) extend the classical Q-learning algorithm by using a neural network to approximate the Q-value function, enabling RL to scale to high-dimensional state spaces. The Bellman equation serves as the foundation for Q-value updates, where the Q-value of a state-action pair is updated based on the immediate reward and the maximum Q-value of the subsequent state. The neural network approximates the Q-value function, and the parameters are updated by minimizing the following loss function:
\begin{equation}
L_i(w_i) = \mathbb{E}_{s,a,r,s'} \left[ \left( r + \gamma \max_{a'} Q(s', a'; w^-) - Q(s, a; w_i) \right)^2 \right]
\end{equation}
Here, $w^-$ are the parameters of a target network updated periodically to stabilize training, $\gamma$ is the discount factor, and $r + \gamma \max_{a'} Q(s', a'; w^-) $ is the target Q-value. DQN stabilizes training through the use of experience replay and target network updates. However, it can struggle in environments with continuous action spaces.

\textbf{Algorithm:}
\begin{itemize}
    \item Initialize the Q-network with random weights and the target network with weights $w^-$.
    \item Initialize the replay memory $D$.
    \item For each episode:
    \begin{itemize}
        \item Observe the initial state $s_0$.
        \item For each step in the episode:
        \begin{itemize}
            \item Select an action using an $\epsilon$-greedy policy.
            \item Execute the action, observe reward $r$ and next state $s'$.
            \item Store the transition $(s_t, a_t, r, s_{t+1})$ in memory $D$.
        \end{itemize}
        \item Sample a random minibatch of transitions from $D$.
        \item Compute the target Q-value:
        \begin{equation*}
            y = r + \gamma \max_{a'} Q(s', a'; w^-)
        \end{equation*}
        
        \item Update the Q-network by minimizing the loss:
        \begin{equation*}
        L_i(w_i) = (y - Q(s_t, a_t; w_i))^2
        \end{equation*}
        \item Periodically update the target network weights: $w^- \leftarrow w_i$.
    \end{itemize}
    \item Repeat until convergence.
\end{itemize}

\bibliographystyle{elsarticle-harv} 
\bibliography{references}




\end{document}